\documentstyle[preprint,aps,eqsecnum]{revtex}

\newcommand{\be}{\begin{equation}}
\newcommand{\ee}{\end{equation}}
\newcommand{\n}[1]{\label{#1}}
\newcommand{\ind}[1]{\mbox{\tiny{#1}}}

\def\bbox{{\,\lower0.9pt\vbox{\hrule \hbox{\vrule height 0.2 cm
\hskip 0.2 cm \vrule  height 0.2 cm}\hrule}\,}}

\begin{document}
\draft \tightenlines

\title{Mining Energy from a Black Hole by Strings}

\author{V.P. Frolov}

\address
{Theoretical Physics Institute, Department of Physics, University
of Alberta, \\ Edmonton, Canada T6G 2J1 \\E-mail:
frolov@phys.ualberta.ca}

\author{and}

\author{D.V. Fursaev}

\address
{Joint Institute for Nuclear Research,
Bogoliubov Laboratory of Theoretical Physics,\\
 141 980 Dubna, Russia\\
E-mail: fursaev@thsun1.jinr.ru}

\maketitle

\begin{abstract}
We discuss how cosmic strings can be used to mine energy from
black holes. A string attached to the black hole gives rise to an
additional channel for the energy release. It is demonstrated that
when a  string crosses the event horizon, its transverse degrees
of freedom are thermally excited and thermal string perturbations
propagate along the string to infinity. The internal metric
induced on the 2D worldsheet of the static string crossing the
horizon describes a 2D black hole. For this reason thermal
radiation of string excitations propagating along the string can
be interpreted as Hawking radiation of the 2D black hole. It is
shown that the  rate of energy emission through the string channel
is of the  same order of magnitude as the bulk radiation of the
black hole.  Thus, for $N_s$ strings attached to the black hole
the efficiency of string channels is increased by factor $N_s$. We
discuss restrictions on $N_s$ which exist because of the finite
thickness of strings, the gravitational backreaction and quantum
fluctuations. Our conclusion is that the energy emission rate by
strings can be increased as compared to the standard emission in
the bulk by the factor $10^3$ for GUT strings and up to the factor
$10^{31}$ for electroweak strings.

\vskip 1truecm

\today

\vskip 1truecm

\end{abstract}
 \vfill\eject

\vfill\eject

\section{Introduction.}

A black hole behaves like a heated body. A black hole of mass $M$ emits
thermal radiation with temperature $(8\pi M)^{-1}$. But there exists a
special property which singles out  a black hole as a thermodynamical
object. While a size of a usual radiating body can be arbitrary, the
size of the black hole is determined by the same parameter $M$ as its
temperature. For given temperature $T$ the rate of radiation of the
heated body, which is proportional to $T^4 R^2$, can be made arbitrary
large by making its size $R$ bigger and bigger. This is not true for a
black hole. The rate of energy emission by the black hole is determined
only by its mass and is proportional to $M^{-2}\sim T^2$.

Can one increase the rate of radiation of the black hole by an
external influence?  Unruh and Wald  \cite{UnWa:82,UnWa:83a}
demonstrated that in principle this is possible. They proposed a
gedanken experiment when a small box with  ideally conducting
(mirror-like) boundaries was taken close to  black hole surface, opened
there, closed again,  and taken back to  infinity. They demonstrated
that in  this process energy can be  extracted from the black hole in
the form of thermal radiation.  By repeating this process, one  can
mine additional energy from a black hole and effectively increase its
emission.  Unruh and Wald conjectured that the highest
possible rate of energy  extraction from a black hole can reach
$c^5/G\approx 3.6\times 10^{59}$erg/s \ \ \cite{UnWa:83b}.

One can describe this process of energy mining in slightly different
terms.  An effective surface area of the black hole is determined by a
position of the potential barrier. This potential barrier does not allow
quanta of the thermal atmosphere of the black hole located close to the
horizon to escape to the infinity. One can describe the effect of
energy mining,  at least in a quasi-stationary regime, by saying that a
device  which is used for this purpose simply increases the probability
of the  penetration   through the potential barrier of  modes in some
frequency range.  As the result, the effective radiative surface area
increases.

In this paper we would like to propose a new mechanism of energy
mining from black holes. Namely we propose to  use for this
purpose {\em cosmic strings}. We demonstrate that this mechanism
is very efficient. In many aspects a `device' for energy mining by
means of strings seems to be much more `realistic' than
Unruh-Wald's gedanken device using idealized boxes.

\section{Mechanism of Energy Mining}

Different aspects of the black-hole energy mining by strings are
discussed in other Sections. Here we describe the main idea and give
some order of magnitude estimations of the efficiency of this process.

Interaction of cosmological defects (cosmic strings, domain walls) with
black holes was discussed in \cite{FrHeLa:96,ChFrLa:98,ChFrLa:99}. In
particular it was shown that there exist stationary configurations of
such defects  which spread from infinity to a black hole and enter  the
black hole horizon.

To describe this situation in a more general set-up let us first assume
that the spacetime has $D$-dimensions and it contains a $D$-dimensional
black hole. We call it a {\em bulk} black hole. Consider a
$d$-dimensional world surface $\Sigma$ representing a topological
defect in this $D$-dimensional spacetime. We call it a {\em brane}. An
internal geometry $\gamma$ on the brane  is generated by its embedding
in the $D$-dimensional bulk spacetime. If the static surface $\Sigma$
crosses the event horizon of the static bulk black hole, the induced
geometry $\gamma$  itself  describes a $d$-dimensional black hole. It
follows from a simple fact that a Killing vector of the bulk space is
tangent to $\Sigma$ and thus it is the Killing vector of the
induced mertic $\gamma$. Hence the surface where $\Sigma$ crosses the
horizon of the bulk black hole is the Killing horizon of the
$d$-dimensional black hole on the brane.

It can be shown that transverse perturbations of  test branes can be
described as a set of massless scalar fields propagating in the
background metric $\gamma$. In the presence of the $d$-dimensional
black hole these degrees of freedom are thermally excited and thermal
radiation of brane excitations  emitted by $d$-dimensional black hole
propagates along the brane to infinity.  Thus such a brane  serves as
an additional channel  and can be used for the energy mining  from the
bulk black hole.

Let us estimate the efficiency of this new additional `channel'. The
required exact calculations are given later (Section~3). Now we
just make an order of magnitude estimation and explain why the mechanism
of energy extraction from black holes by means of branes can be
effective.

The energy flux from a black hole is
\begin{equation}\label{i.1}
\dot{E}\equiv {dE\over dt}={1\over 2\pi}\int_{0}^{\infty}\,
{d\omega \, \omega\,\Gamma_{\omega}\over \exp(8\pi M\omega)-1}\, ,
\end{equation}
where $\Gamma_{\omega}$ is the transition probability for a mode
of energy $\omega$ to penetrate the potential barrier and to reach
the infinity.   This formula is valid for any number of spacetime
dimensions.  Let us apply it to the radiation of the
$d$-dimensional black hole on the brane. Notice that the internal
$d$-dimensional black-hole geometry on the static brane crossing
the  bulk black hole is characterized by  only one parameter $M$,
related to the mass of the bulk black hole, or what is equivalent,
by its gravitational radius\footnote{The parameter $M$ is
proportional to the gravitational radius. $M$ coincides with the
black hole mass $M_{BH}$ in four dimensions. In higher dimensions
$D$ the relation is more complicated $M_{BH}\sim M^{D-3}$, see
\cite{MP}.}. Thus, the equation for a massless field propagating
in this background can be easily rewritten in the dimensionless
form where the dependence on $M$ enters only through dimensionless
quantity $\varpi=\omega/T_{BH}=8\pi M\omega$.\footnote{We use
units in which $G=c=\hbar=1$.} Because the factor
$\Gamma_{\omega}$ is dimensionless and depends on $\varpi$ one has
\be\n{i.2} \dot{E}={J_d\over 128\pi^3 M^2} \, ,\hspace{0.5cm}
J_d=\int_{0}^{\infty}\, {d\varpi \, \varpi\, \Gamma(\varpi)\over
\exp(\varpi)-1}\, . \ee This observation shows that the energy
density emitted by a $d$-dimensional black hole is always
proportional to $M^{-2}$, and  the dimensionality $d$ appears only
in a dimensionless coefficient $J_d$. This important simple
property was recently discussed by  Emparan, Horowitz and Myers
\cite{EHM} who examined Hawking radiation for black holes on a
brane in a world with large extra dimensions.

Thus, a  static $d$-dimensional topological defect attached to the bulk
black hole gives rise to additional radiation with the flux
$J_d M^{-2}$.  The rate of energy emission through  the new channel has
the same order of magnitude as the emission rate of the bulk black
hole. The more branes are attached to the bulk black hole the higher
efficiency of the energy mining. One can expect that  the number of
such branes which can be attached to the bulk black hole without their
mutual intersections is bigger for small $d$.

Hence, returning to the discussion of  the physically most interesting
case of 4-dimensional bulk spacetime one can conclude that the highest
efficiency of the black hole energy mining would be obtained if one
uses cosmic strings ($d=2$).  In the present paper we discuss this idea
and some of its consequences.

\section{Thermal String Excitations}

To discuss quantum radiation of string excitations let us consider
a static string of tension $\mu$ which enters radially a
Schwarzschild black hole of mass $M$ through its north pole. The
other part of  the string (string segment) which enters radially
the black hole through the south pole  is not of interest for the
moment. Denote by $n^{\mu}_R$, $R=1,2$, two  mutually orthogonal
normal vectors to the string world-sheet $\Sigma$.  Small
perturbation of the string $\delta x^{\mu}$ can be decomposed  in
terms of \be\n{2a.1} \Phi_R=\delta x^{\mu}\, n^{\mu}_R \ee and a
component along the world-sheet, which is a pure gauge.  By
linearizing the Nambu-Goto  action\footnote{ Here the constant
$\mu$ is the tension of the string measured in units $G=c=1$. We
keep in mind that it corresponds to the dimensionless combination
$G\mu/c^2$.} \be\n{2a.2} I=-\mu\int d^2 \zeta \sqrt{\gamma}\, ,
\ee ($\gamma_{AB}$ is the induced metric on $\Sigma$) one obtains
the following quadratic action for string perturbations
\be\n{2a.3} I[\Phi]=-\mu\, \int d^2 \zeta \sqrt{\gamma}\, \left[
\gamma^{AB} \Phi_{P,A}\Phi_{R,B}+ {1\over 2}\, R\, \Phi_P\,
\Phi_R\right]\, \delta^{PR}\, , \ee where $R$ is a scalar
curvature of the 2-dimensional induced metric $\gamma$. We call
these string excitations {\em stringons}. The induced metric for
the unperturbed string is \be\n{2.4} d\gamma^2= \gamma_{AB}\,
d\zeta^A\, d\zeta^B= -F\, dt^2 + F^{-1}\, dr^2\, ,\hspace{0.5cm}
F=1-{2M\over r}\, , \ee $\zeta^{A}=(\zeta^0,\zeta^1)=(t,r)$, and
$R=4M/r^3$.  By redefining the fields $\Phi_R$ \be\n{2a.5}
\varphi_R = \sqrt{2\mu}\, \Phi_R\, , \ee we can write (\ref{2a.3})
as a sum of two identical actions each being of the form
\be\n{2a.6} I[\varphi]=-{1\over 2}\, \int d^2 \zeta
\sqrt{\gamma}\, \left[ (\nabla\varphi)^2+{1\over 2}\, R
\varphi^2\right]\, . \ee This is nothing but a massless
non-minimally coupled ($\xi=-1/2$) scalar field in the spacetime
of a two-dimensional Schwarzschild black hole of mass $M$.

The field equation
\be\n{2a.7}
\left[\Box -{1\over 2}R\right]\varphi
=0\, ,
\ee
in the tortoise coordinates $dr_*=dr/F$ takes the form
\be\n{2a.8}
\left[ -{\partial^2\over \partial t^2}+{\partial^2\over
\partial r_*^2} +U\right] \varphi =0\, ,\hspace{0.5cm} U=-{2M\over
r^3}(1-{2M\over r})\, ,
\ee
which is identical to the equation for $s$-mode of a  4-dimensional
massless scalar field in the Schwarzschild geometry.  Since stringons
have two independent states of polarization, any string segment
attached to the bulk black hole brings with it 2 additional channels of
emission. The rate of emission through each of these channels is
identical to the $s$-mode contribution to the rate of emission  for a
scalar massless field in the bulk space.

To estimate this contribution one can use the following  approximation
for $\Gamma$ at low frequency obtained by Starobinsky \cite{St:73}
and Page \cite{Page:76}
\footnote{Notice that this result is 10 times smaller than the rate
emission calculated in the absence of the potential barrier which can
be obtained by using  Polyakov effective action \cite{FrVi:81}.}
\be
\Gamma_{\omega} = 16\, \omega^2 \, M^2\, .
\ee
The corresponding rate of emission is
\be\n{2a.10}
\dot{E}^{\ind{SP}}\equiv {16\, M^2\over 2\pi}\,
\int_{0}^{\infty}\, \, {d\omega \,
\omega^3 \over \exp(8\pi M\omega)-1}= {1\over 7680 \pi\,M^2}\, .
\ee
Numerical calculations give for the exact value of the $s$-mode
contribution the following result
\be
\dot{E}^{\ind{s-mode}}= 1.62 \dot{E}^{\ind{SP}}\, .
\ee

To compare this result with the rate of bulk emission of the black
hole we remind that the approximate expression can be obtained by
using DeWitt's anzats \cite{DeWitt,bhbook} \be\n{2a.9a}
\Gamma_{\omega} \approx 27\, M^2\, \omega^2 \, . \ee The emission
rate in this approximation is \be\n{2a.11}
\dot{E}^{\ind{DeWitt}}\equiv {27 M^2\over 2\pi}\,
\int_{0}^{\infty}\, \, {d\omega \, \omega^2 \over \exp(8\pi
M\omega)-1}={9\over 40960\pi M^2}\, . \ee Numerical calculations
\cite{Elst:83,Simp:86} shows that this result is very accurate. In
fact the exact value $\dot{E}^{\ind{scalar}}_{\ind{bulk}}$ of the
bulk rate of emmision is
\be
\dot{E}^{\ind{scalar}}_{\ind{bulk}}=1.06\, \dot{E}^{\ind{DeWitt}}
\ee By using these results we can write \be\n{2a.12}
\dot{E}^{\ind{s-mode}}=0.91 \dot{E}^{\ind{scalar}}_{\ind{bulk}}\,
.
\ee
In other words, the additional rate of quantum radiation
$\dot{E}^{\ind{string segment}}$ through each segment of the
string attached to the bulk black hole is 1.8 times bigger that
the contribution of the scalar massless field to the emission rate
of the bulk black hole\footnote{The fact that the energy flux from
a black hole in string excitations is approximately that of field
modes in the bulk was pointed out earlier by Lawrence and Martinec
\cite{LM}.}. We took into account that there are two states of
polarization of the string excitations (stringons).

Above we assumed that the bulk  black hole radiated a massless scalar
field. In fact a realistic black hole emits all possible species of
particles which can be emitted at the
given Hawking temperature.
Black holes of mass greater than $10^{17}$g effectively emit only
massless particles (neutrinos, photons, and gravitons). Using the
results obtained by Page \cite{Page:76,Page:76a} one has
\be
\dot{E}^{\ind{bulk BH}}= A_{m=0}\, \dot{E}^{\ind{scalar
field}}_{\ind{bulk}}\, ,
\ee
where
\be\n{a3.17}
A_{m=0}= 0.54\, h(1/2)+ 0.22\, h(1)\,+ 0.03\, h(2)\, ,
\ee
and $h(s)$ is the number of distinct polarizations of spin-$s$
particles. For photon, graviton and
3 sorts of massless neutrino $A_{m=0}= 3.74$.
Thus for a black hole of the
mass $M>10^{17}$g which emits only massless particles one has
\be
\dot{E}^{\ind{string segment}}\approx {1\over 2} \dot{E}^{\ind{bulk
BH}}\, .
\ee
For smaller black holes the right-hand side of this relation contains
an additional factor $A_{m=0}/A_{\ind{total}}$, where $A_{\ind{total}}$
is the factor $A$, (\ref{a3.17}),  calculated for all particles which
can be effectively emitted at the temperature $(8\pi M)^{-1}$.

\section{Efficiency of Black Hole Mining by Strings}

Let us discuss now how many strings can be attached to the black hole.
There are several kinds of restrictions on the number of strings. The
most important of them are  effects connected with the string's
thickness, and effects connected with the backreaction.

Before discussing these effects let us make the following remark.
A string heated to the temperature $T$ higher than the energy of the
corresponding phase transition $\sim \mu^{1/2}$ becomes unstable and
can decay. One can expect this kind of instability if the string is
attached to the black hole with Hawking temperature $(8\pi M)^{-1}$
higher than $\mu^{1/2}$. The thickness of the string with tension
$\mu$ is proportional to $\mu^{-1/2}$. Hence one can expect that this
effect of thermal instability is not dangerous if the thickness of the
string is much smaller than the gravitational radius of the black hole.
We shall always assume that this condition is satisfied.

Let us consider now the effects connected  with the thickness of
strings\footnote{ A detailed picture of how a Nielsen-Olesen
string of a finite width interacts with a black  hole is discussed
in \cite{DGT:92}. The effects connected with the string thickness
are also important for high-frequency waves propagating along the
string when the wave length becomes comparable  with $\mu^{-1/2}$.
}. Denote by $N_s$ the number of string segments attached to the
black hole of mass $M$. Then the total surface area of string
cross-section is proportional to $\pi N_s\mu^{-1}$. If the strings
do not touch each other this quantity must be smaller than the
surface area of the black hole, $16\pi M^2$. This condition gives
the following restriction on $N_s$ \be\n{2a.14} N_s <16 M^2\,\mu\,
. \ee

Second kind of restrictions on $N_s$ is connected with the
backreaction effect.  The gravitational field of a string creates
an angle  deficit $\alpha= 4\mu$, that is the length of a unit
circle   around the string becomes $\beta=2\pi(1-\alpha)$. It can
be shown  by using Gauss-Bonnet theorem that $N_s$ string segments
of the equal tension $\mu$ which enter the  black hole horizon
change the solid angle around the black hole from $4\pi$ to
$2\pi(2-N_s\alpha)$. This means that we have to consider only
situations when \be\n{2a.13} a\equiv 2\, \mu\, N_s \ll 1\, . \ee

For a certain mass  $M=M^*=(a/32)^{1/2}/\mu$ the both restrictions give
the same  number $N_s$. For $M<M^*$ ($M>M^*$) restriction on $N_s$
follows from relation (\ref{2a.14}) (from relation (\ref{2a.13})).  To
give estimations we consider two values of the string tensions,
$\mu_{GUT}$ and $\mu_{EW}$,   which correspond to the energy scale of
the Grand Unified  ($10^{16}$ GeV) and Electroweak ($100$ GeV) phase
transitions,  respectively. One has
\be\n{2a.15}
\mu_{GUT}=10^{-6}\, ,\hspace{0.5cm}\mu_{EW}=10^{-34}\, .
\ee
For estimations we also choose $a=1/100$. Then we have
\be\n{2a.16}
M^*_{GUT}= 1.77\times 10^4 m_{Pl}\approx 0.2 \mbox{g}\, ,\hspace{0.5cm}
M^*_{EW}= 1.77\times 10^{32} m_{Pl}\approx 2\times 10^{27} \mbox{g}\, .
\ee
It is interesting that the mass $M^*_{EW}$ is close to the Earth's mass
$10^{28}$g.

For GUT strings $M^*_{GUT}$ is so small that for all black holes
which are not at the stage of their explosion one has $M>M ^*_{GUT}$ and
hence the upper limit on $N_s$ is
\be\n{2a.17}
N_s^{GUT}\sim 5\times 10^3\, .
\ee
For $N_s^{EW}$ the restrictions are
\be\n{2a.18}
N_s^{EW} \sim 5\times 10^{31}\, ,\mbox{\ for }\ M>M^*_{EW}\, ,
\ee
and
\be\n{2a.19}
N_s^{EW} \sim 5\times 10^{31}\left({M\over M^*_{EW}}\right)^2\, ,
\mbox{\ for }\ M<M^*_{EW}\, .
\ee
One can conclude that even with GUT strings mining energy
from black holes can be quite efficient. For electro-weak or
lighter strings the effect of amplification of the rate of quantum
radiation is very high.

Let us estimate now  the total energy rate of emission
through the attached strings. For $M>M^*$ the number of attached
string segments is $N_s^*=a/(2\mu)$ and
\be
\dot{E}^{\ind{string}}_{\ind{total}}={B\, N_s^*\over M^2}\, ,
\ee
where $B\approx 1/(2400\pi)$. $\dot{E}^{\ind{string}}_{\ind{total}}$
grows when $M$ becomes smaller until it reaches  $M^*$. After
that its value remains constant. Denote this maximum value of the energy
rate by $\dot{E}^*$, then one has
\be\n{4.9}
\dot{E}^*\approx 2.5\times 10^{-3}\, \mu \ \dot{E}^{\ind{max}}\, ,
\ee
where $\dot{E}^{\ind{max}}=c^5/G$ is the highest rate of energy
extraction discussed in the Introduction.

\section{Thermal Fluctuations of Strings in Black Hole Geometry}

Thermal quantum fluctuations of the string near the black hole
horizon can change effectively the string cross section. If this
effect would be large we had to reconsider estimations of the
previous Section. Let us analize this effect in more detail. The
effective width $\Delta x$ generated by quantum fluctiations of
the string can be defined as
\begin{equation}\label{a.10}
(\Delta x)^2=\langle \delta x^\mu \delta x_\mu \rangle~~~,
\end{equation}
where $\delta x^\mu$ are transverse perturbations of the string.
The quantum average can be defined according to (\ref{2a.1}) and
(\ref{2a.5}) as
\begin{equation}\label{a.11}
\langle \delta x^\mu \delta x_\mu \rangle= \sum_R \langle
\hat{\Phi}^2_R \rangle={1 \over \mu} \langle \hat{\varphi}^2
\rangle~~~,
\end{equation}
where $\varphi$ is a two-dimensional quantum scalar field
described  by equation (\ref{2a.7}). To estimate the effective
width of the string fluctuations we have to find the correlator
$\langle \hat{\varphi}^2 \rangle$ on the horizon of the 2D black
hole. We will do it by calculating explicitly the Green function
$G(p,p')$ of the field $\varphi$ in the Euclidean theory.

The 2-dimensional Euclidean black hole metric can be written as
\be\n{a.1} d\gamma_E^2=4M^2 d\tilde{\gamma}^2\, , \hspace{0.5cm}
d\tilde{\gamma}^2= (1-x^{-1})\, d\tau^2 +(1-x^{-1})^{-1}\, dx^2\,
, \ee where $x=r/2M$ and $\tau$ is periodic with the period
$2\pi/\kappa=4\pi$.  Here $\kappa=1/2$ is the surface gravity of
the horizon in the metric  $d\tilde{\gamma}^2$. Near the horizon
the metric $d\tilde{\gamma}^2$  takes the Rindler form \be\n{a.1a}
d\tilde{\gamma}^2= {l^2\over 4}\, d\tau^2 +dl^2\, , \ee where $l$
is the dimensionless proper distance.

To calculate thermal fluctuations at the horizon of a string
attached to a black hole it is sufficient to know the Green
function $G(p,p_0)$ with one point at the horizon $x=x_0=1$. Such
a Green function does not depend on $\tau$ and can be written as
$G(x,x_0)$. The Green function equation \be\n{a.2} \Box \,
G(p,p')=-\delta(p,p')\, , \ee takes the form \be\n{a.3} \left[{d
\over d x}\left( (1-{1\over x})\, {d \over d x}\right) -{1\over
x^3}\right] G(x,x_0)=-{\delta(x-x_0)\over 4\pi}\, . \ee A general
solution of this equation  is \be G(x,x_0)=C x +{1\over 4\pi} x
\ln{x\over x-1}\, . \ee To single out a solution we impose a
boundary condition $G(L,x_0)=0$, that is we suppose that a
position of the string is fixed at the point $x=L$. Then we have
\be G(x,x_0)={1\over 4\pi} x \left[ \ln{x\over x-1} - \ln{L\over
L-1} \right]\, . \ee

It is not difficult to see that if the point $p$  is approaching
$p_0$  then $(x-1)=(\Delta l)^2/(4M)^2$ where $\Delta l$ is the
proper distance between the two points. The distance $\Delta l$
can be considered as a  regularization parameter, a cutoff which
enables one to avoid short-distance or ultraviolet divergences  in
the average of the operator $\hat{\varphi}^2$. Thus, the
regularized correlator can be written as \be\n{a.12} \langle
\hat{\varphi}^2\rangle ={1\over 4\pi} \left[ \ln\left( 4M\over
\Delta l\right)^2-\ln {L\over L-1}\right]\, . \ee In the
considered situation the value of ultraviolet cutoff is related to
the thickness of the string. The high-frequency waves with the
wave-length comparable to or smaller than the string width  cannot
propagate along the string, or at least their propagation should
be different from the propagation of the  low-frequency waves.
Hence, it is natural to relate $\Delta l$ with the  string width
$1/\sqrt{\mu}$.

The quantity  $b\equiv 4M\sqrt{\mu}$ is large for the
macroscopical black holes. For instance, for a solar mass black
holes $b=10^{35}$ for GUT strings and $b=10^{21}$ for Electroweak
strings. However, $b$ appears in the right hand side of
(\ref{a.12}) under the logarithm. Because of this additional
logarithm the effective width of the string $\Delta x$ related to
quantum fluctuations can be only order or two greater than the
string width $1/\sqrt{\mu}$. This factor can be easily taken into
account in  the estimations of the previous Section of the upper
number of strings  which can be attached to the black hole but
this does not change essentially the main results.

\section{Polyhedral String Configurations}

Our device for energy extraction consists of  a black hole and $N_s$
segments\footnote{If a black hole captures an infinite cosmic string, 2
new string segments would be attached to the black hole. For this
reason $N_s$ must be an even number.} of  strings attached to the black
hole which are located along radial directions.

Till now we did not discuss gravitational back reaction and
possible mutual  interaction  of strings attached to the black
hole. String interactions can arise for different reasons.
Consider two parallel Nielsen-Olesen strings separated by distance
$l$ greater than $\mu^{-1}$.  Strictly speaking the fields do not
vanish outside the radius $\mu^{-1}$, but become exponentially
small. Because of the non-linearity of the field equations one can
expect a force $\sim\exp(-l\mu)$ between two strings. Quantum
effects also results in the force between 2 parallel strings
\cite{Bordag}.  In a general case in the presence of these and
mutual gravitational interaction forces (proportional to $\mu^2$)
a system of strings is not static. One can expect that when the
distances between strings are large in comparison with their width
these forces are negligible.

Quite remarkably, there is a special case when one can guarantee
that the forces acting on the strings identically vanish because
of the symmetry \cite{FF:2001}.  It happens when there exists a
symmetry transformation (rotation) which transforms the system
(the black hole with attached strings) into itself. The force
acting on the string (which is always orthogonal to the string)
must vanish since it must remains invariant under this rotation.
The string configurations which respect this symmetry are
discussed in detail in \cite{FF:2001}. We just mention here that
for these special (polyhedral) configurations strings are directed
along radii and coincide with the symmetry axes of regular
polyhedra. Therefore, there exist only few types of string
configurations.  The number of strings corresponding to a
tetrahedron, octahedron and icosahedron  is 14, 26 and 62,
respectively\footnote{The number of strings differs from the order
of the symmetry group (12, 24, 60) of the corresponding solids.
This difference is discussed in \cite{FF:2001}.}. There is also a
so called "double pyramid configuration" when an even number of
strings goes inside the black hole in the equatorial plane at the
equal angles between two neighbor strings and the two strings are
attached at the south and the north poles. The number of stings in
this configuration can be arbitrary. All strings  have equal
tension.\footnote{In a more general case strings corresponding to
three different types of the  symmetry axes can have different
tensions.}  For an arbitrary  string tension the symmetry
guarantees that such configurations are exact solutions of the
Einstein equations and that outside the strings the metric is
locally isometric to the Schwarzschild metric \footnote{If the
force between two strings in the Schwarzschild geometry is
repulsive  polyhedral string configurations are stable. The
problem of stability of polyhedral strings requires further
investigation.}.

Returning to the problem of the energy mining from  black holes we see
that the icosahedron configuration  gives about $10^2$  independent
channels for the energy mining. For the tetrahedron and octahedron
configurations this number is less. The maximal number of strings is
possible for the double pyramid configuration.  As we discussed above
there is a  restriction on the total number $N_s$ of strings in this
configuration determined by value of the solid angle around the black
hole, $a=2\mu N_s \ll 1$. The other restriction is the upper bound on
the number of "equatorial" strings of the width $\mu^{-1/2}$ which can
be attached to the black hole of the horizon radius $2M$
\begin{equation}\label{1.17}
N_s\simeq 4\pi M \sqrt{\mu}~~~.
\end{equation}
The both restrictions result to the same upper bound $N_s$ at
$M=M^{*}=a/(8\pi)\mu^{-3/2}$.  For $M<M^*$ ($M>M^*$) restriction
on $N_s$  follows from relation (\ref{1.17}) (from relation
(\ref{2a.13})).  We can again give estimations for two values
$\mu_{GUT}$ and $\mu_{EW}$,   which correspond to Grand Unified
and Electroweak scales,  respectively.  By choosing as before
$a=1/100$ we have \be\n{1.18} M^*_{GUT}\sim 4 \mbox{g}\,
,\hspace{0.5cm} M^*_{EW}\sim  4\times 10^{42} \mbox{g}\, . \ee For
a black hole with the mass  $M> 10^{17}\mbox{g}$ we get the same
upper bound (\ref{2a.17}) for GUT strings ($N_s \sim 10^3$). For
Electroweak strings the restrictions are
\be\n{1.19} N_s^{EW} \sim
5\times 10^{31}\, ,\mbox{\ for }\ M>M^*_{EW}\, , \ee and
\be\n{1.20} N_s^{EW} \sim 5\times 10^{31}{M\over M^{*}_{EW}}\, ,
\mbox{\ for }\ M<M^*_{EW}\, . \ee
Bound (\ref{1.19}) coincides
with (\ref{2a.18}). For a stellar mass black hole $M\sim 10^{34}$g
one has to use (\ref{1.20}) which differs from (\ref{2a.19}). The
number of strings in this case is $N_s\sim 10^{23}$.

We have thus demonstrated  that there are black hole solutions with
many strings where the backreaction effect is taken into account
explicitly and where the upper bound on the number of strings
either coincides with our general estimations or somewhat different
but still very high.

It should be emphasized that one can expect  the existence of wider
variety of static solutions of Einstein equations describing a static black
hole with attached radial cosmic strings than polyhedral configurations.
We discuss this problem elsewhere \cite{FF:2001}.

\section{Discussion}

We demonstrated that mining energy from black holes by cosmic
string can be very effective. In principle,  a black hole with
attached cosmic strings can be used as an energy generator. To
make it `technically' simpler, one may assume that strings in the
black hole exterior have end points with monopoles which are
attached to a spherical construction with the size  of a few
gravitational radius. Thermal energy released in this construction
can be transformed into useful work. If we use the relation
(\ref{4.9}) to estimate  the energy generated by this device  we
get for the Electroweak strings the result $\dot{E}^*\sim 9\times
10^{22}$erg/sec. This is  one order of magnitude smaller that the
total solar radiation through the Earth surface which is
approximately $1.7\times 10^{24}$erg/sec. Such a device is able to
produce energy $\dot{E}^*$ during period of time $T\sim
Mc^2/\dot{E}^*$. For example for a black hole of the same mass as
the mass of Earth, $\sim 6\times 10^{27}$g, the time $T$ is
approximately $10^{26}$sec, that is much greater than the lifetime
of the Universe.

In our discussion we assumed that black holes are non-rotating.  One
can expect that for rotating black holes the estimations obtained above
are also valid. Stationary configurations of test strings near a
rotating black hole were discussed in \cite{FrHeLa:96}. According to
the uniqueness theorem proved there, the only stationary strings which
enter a stationary rotating black hole have worldsheets spread by a
timelike at infinity Killing vector and a principle null ray of the
Kerr geometry. Such a string belongs to a cone $\theta=$const.
Different non-intersecting strings which belong to the same cone can be
obtained by a rigid rotation at some angle $\phi$. Strings which belong
to different cones do not intersect one another. Thus such strings form
a two-parameter family which is parameterized by angles $(\theta,
\phi)$, the coordinates on the infinite red-shift surface of the points
where a string crosses this surface. As it was demonstrated in
\cite{FrHeLa:96} the internal geometry on such a stationary string
surfaces is again 2D black-hole geometry so that there is thermal flux
of stringons  along the strings.

A black hole and a cosmic string are relativistic objects with well
known (at least theoretically) properties. The assumptions we have
made to study the  strings attached to the black hole can be easily
controlled. This makes this system much simpler for analysis than
Unruh-Wald gedanken experiment with  mirror boxes
\cite{UnWa:82,UnWa:83a,UnWa:83b}. It is  interesting that quite general
arguments allows one to obtain the relation (\ref{4.9}) for the maximal
rate of radiation in the string--black-hole system, which besides the
universal `maximal' rate $c^5/G$ contains only one additional factor
$\mu$, the string tension.

Finally, we want to mention an another problem which is not directly
related to strings in four dimensions but where the above results may
be of some interest as well.  There is an analog of string-like defects
in higher dimensional  gravity theories. As was shown in \cite{GS:00},
it is possible to obtain a four-dimensional gravity  from a
six-dimensional gravity localized on a string-like defect. This
scenario may be an alternative to localizing gravity on branes. Our
arguments then can be used to show that the Hawking emission rate of a
six-dimensional black hole along a string-like defect (i.e., into a
four-dimensional world) would be comparable to its radiation in the
bulk.   This is similar to what one can say about radiation of a black
hole along the brane \cite{EHM}.

\noindent
\section*{Acknowledgments}

\indent This work was
partially supported  by the Natural
Sciences and Engineering
Research Council of Canada, by the NATO Collaborative Linkage Grant
CLG.976417. D.F. was also supported by RFBR grant N99-02-18146.

\end{document}